\documentclass[12pt, a4paper]{article}
\pdfoutput=1
\usepackage{graphicx}
\usepackage{amsmath}
\usepackage{multirow}

\begin{document}
\title{A generalised mean-field approximation for the Deffuant opinion dynamics model on networks.}

\author{Susan C. Fennell\footnote{MACSI, Department of Mathematics and Statistics, University of Limerick, Limerick, Ireland}\hspace{0.4cm}
Kevin Burke\footnotemark[1]\hspace{0.4cm}
Michael Quayle\footnote{Department of Psychology, University of Limerick, Limerick, Ireland} \footnote{Department of Psychology, University of KwaZulu-Natal, Pietermaritzburg, KwaZulu-Natal, South Africa}\hspace{0.4cm}
James P. Gleeson\footnotemark[1]
 }

\date{}
\maketitle
\begin{abstract}

When the interactions of agents on a network are assumed to follow the  Deffuant opinion dynamics model, the outcomes are known to  depend on the structure of the underlying network.
This behavior cannot be captured by existing mean-field approximations for the Deffuant model. In this paper,  a generalised mean-field approximation is derived that accounts for the effects of network topology on Deffuant dynamics through the degree distribution or community structure of the network.  The accuracy of the approximation is examined by comparison  with large-scale Monte Carlo simulations on both synthetic and real-world networks.

\end{abstract}

\section{Introduction}

Mean-field approximations can be used to gain insight into the behaviour of complex dynamical systems at a fraction of the computational cost associated with running large-scale Monte Carlo simulations. Originally developed to study phase transitions in statistical physics, mean-field (MF) theory is now used in various areas, for example epidemic modelling \cite{PastorSatorras2001} and neural networks \cite{Bick2020}, but our focus is its use in the study of models of opinion formation and consensus \cite{Vazquez2004, Carro2016,Chen2018,Krapivsky2003,Castellano2000,Vilone2002,Ben-Naim2003}. Mean-field approximations are particularly useful for studying dynamics that take place on the nodes of a (large) network.  In models of opinion dynamics, for example, each node in a network represents an agent who holds an opinion, and the model dictates how the agents interact to change their opinions over time.  By making assumptions about the network structure and dynamical correlations \cite{Porter2016}, a set of equations for the  time-dependent proportion of agents with a given opinion can be derived. The number of such equations is typically much smaller than the system size and so numerical integration of those equations is more efficient than Monte Carlo simulations of the entire system. In some cases it is possible to identify important parameters in the system through a mathematical analysis of the MF equations. 
In the Axelrod model, for example, numerical integration of the MF equations reveals a phase transition between an ordered phase in which all agents hold the same set of opinions and a disordered phase in which the network is separated into non-interacting clusters of agents with the same opinions \cite{Castellano2000}. The precise location of this phase transition for the one dimensional case can be obtained though a mathematical analysis of the MF equations \cite{Vilone2002}.

The Deffuant model is a so-called ``bounded confidence'' model of opinion formation, whereby agents are only influenced by those agents whose (scalar) opinions are close to (within a parameter $\epsilon$ of) their own \cite{Deffuant2000}. While the Deffuant model has mostly been studied by Monte Carlo simulations of the dynamics, a MF approximation was developed and analysed in \cite{Ben-Naim2003}, under the assumption of a well-mixed population of agents. Under this assumption, which is also a common assumption in Monte Carlo simulations of the Deffuant model, each agent may interact with any other agent whose opinion is close enough to their own. 
Instead of keeping track of how each individual changes their opinion over time, the MF approach is to focus on the density $P(x,t)$ of opinions $x$ over the population and how this evolves in time. Having derived an evolution equation for $P(x,t)$, the authors of \cite{Ben-Naim2003} numerically integrated the equation for a range of values of the confidence bound $\epsilon$, enabling them to examine the  dependence of the number of opinion clusters at steady state on $\epsilon$.

Notwithstanding its usefulness, note that the approach in  \cite{Ben-Naim2003} makes use of a classical MF assumption: that each agent can (if their opinion allows) interact with any other agent. When we consider the agents as nodes on a fixed network, the well-mixed assumption of \cite{Ben-Naim2003} corresponds to the dynamics taking place on a complete (and infinite) graph, where an agent has every other agent as a network neighbor. However, if (as we wish to do here) the agents are instead assumed to be connected by a fixed social network, so that the choice of potential interaction partners is limited to neighboring nodes on the graph, it was shown in \cite{Meng2018} that
 that the results of running Deffuant dynamics depend on the specific network structure. For example,  on Erd\H{o}s R\'{e}nyi graphs, as on complete graphs, the number of stable opinion clusters increases as the confidence bound $\epsilon$ decreases, while on cycle graphs the confidence bound has no effect - everyone will eventually have the same opinion. The behaviour on lattices \cite{Deffuant2000}, Barab\'{a}si-Albert networks \cite{Stauffer2004a} and Watts Strogatz networks \cite{Gandica2010} also differs from that of complete graphs. Thus, the original MF approximation of \cite{Ben-Naim2003} is insufficient to accurately describe these cases and  a more general MF approximation is needed to describe the way in which Deffuant dynamics are affected by the network structure.

In this paper we first develop a \emph{degree-based MF approximation} that is suitable for configuration-model networks with prescribed degree distribution $q_k$ (and in the limit of infinite network size). To do so, we extend  the density $P(x,t)$ of opinions at time $t$ that was introduced in \cite{Ben-Naim2003} to multiple opinion densities $P_k(x,t)$, one for each distinct degree class $k$ (i.e., group of nodes with the same degree $k$) in the network. By making mean-field assumptions that are common when studying, for example, binary-state dynamics on networks \cite{Porter2016}, we derive a system of coupled equations for the evolution of the various densities $P_k(x,t)$.

We also develop a further generalization of this method, where we consider a partition of the network nodes into discrete classes, with the probability for two nodes being connected on the network depending only on their class labels. This reduces to our first case when the class of a node is its degree, but we show that other useful partitions can also be considered in this way. 
This more general approximation, which we refer to as a \emph{class-based MF approximation}, can be used in cases where the degree distribution alone does not describe the network structure well, for example, in networks with community structure such as stochastic block models.
By comparing with Monte-Carlo simulations, we show that the class-based MF equations yield a good approximation to the Deffuant dynamics and that they accurately predict behavior on networks that the original MF approximation of \cite{Ben-Naim2003} fails to capture due to its well-mixed assumption being violated.

The rest of this paper is organised as follows. In section~\ref{sec:effect_network} we introduce the Deffuant dynamics and give a motivating example where network structure affects the outcome. We derive the degree-based MF equations for the Deffuant model in section~\ref{sec:dbmf} and compare the solution of these equations with the simulation results from section~\ref{sec:effect_network}. In section~\ref{sec:cbmf} we derive the more general set of class-based MF equations. We use these equations to describe the dynamics on networks with community structure in section~\ref{sec:sbm} and on examples of real-world networks in section~\ref{sec:real_networks}. We conclude in section~\ref{sec:conclusions}.

\section{Effect of network structure on Deffuant dynamics}
\label{sec:effect_network}
\subsection{Deffuant Dynamics}
In the Deffuant model of opinion dynamics, each node (or agent) $i$ holds a time-dependent opinion $x_i(n)$ that is a real number on the interval $[0,1]$ which defines the continuous space of opinions. 
We consider an undirected network of $N$ nodes together with a set of initial opinions $x_i(0), \,i \in \{1, \ldots, N\}$.
At each discrete time $n$ an edge $(i,j)$ is chosen at random from the network. Nodes $i$ and $j$ will update their opinions $x_i(n)$ and $x_j(n)$ if their opinion difference $\left| x_i(n)-x_j(n)\right|$ is below a threshold $\epsilon$. 
The update rule is 
\begin{align*}
x_i(n+1) = x_i(n)+\mu (x_j(n) - x_i(n)),\\
x_j(n+1) = x_j(n)+\mu (x_i(n) - x_j(n)),
\end{align*}
where $\mu \in (0,0.5]$ is a parameter that represents how much individuals are willing to change their opinion. If $|x_i(n) - x_j(n)|\geq\epsilon$ then no change occurs at time $n+1$. 
The dynamics eventually lead to clusters of individuals in the opinion space with members of the same cluster having the same opinion in the limit $n\rightarrow \infty$ \cite{Deffuant2000,Ben-Naim2003}.
On complete graphs clusters form based on opinion distance alone, while on other networks note that individuals whose opinions are within the $\epsilon$ threshold of each other may be disconnected on the network; hence, clusters in the latter context are formed both through individuals' opinions \emph{and} their positions on the network.

\subsection{Examples of dynamics on networks with two degree classes\label{sec:twodegree}}

To illustrate the effect that network structure can have on the dynamics, we simulate the Deffuant model on networks from a configuration model with two degree classes. Each network has $N=10^4$ nodes, 10\% of which (the \emph{minority}) have degree $k_{min} = 100$ while the \emph{majority} have degree $k_{maj}$. We consider two cases: (i) $k_{maj}$ = 5 and (ii) $k_{maj}$ = 25. In each simulation a new network is generated along with a set of initial opinions which are drawn from a uniform distribution on $[0,1]$. We carry out $10^4$ simulations with $\epsilon = 0.3$ and $\mu = 0.5$ for each value of $k_{maj}$. The time $t$ is measured as the number of timesteps divided by $N/2$ so that in one unit of time every node will update their opinion once on average.
We consider the dynamics to have converged when the nodes are separated into non-interacting clusters with opinion difference between any pair of individuals in a cluster less than 0.02, as in \cite{Meng2018}.

\begin{figure}[!t]
\centering
\includegraphics{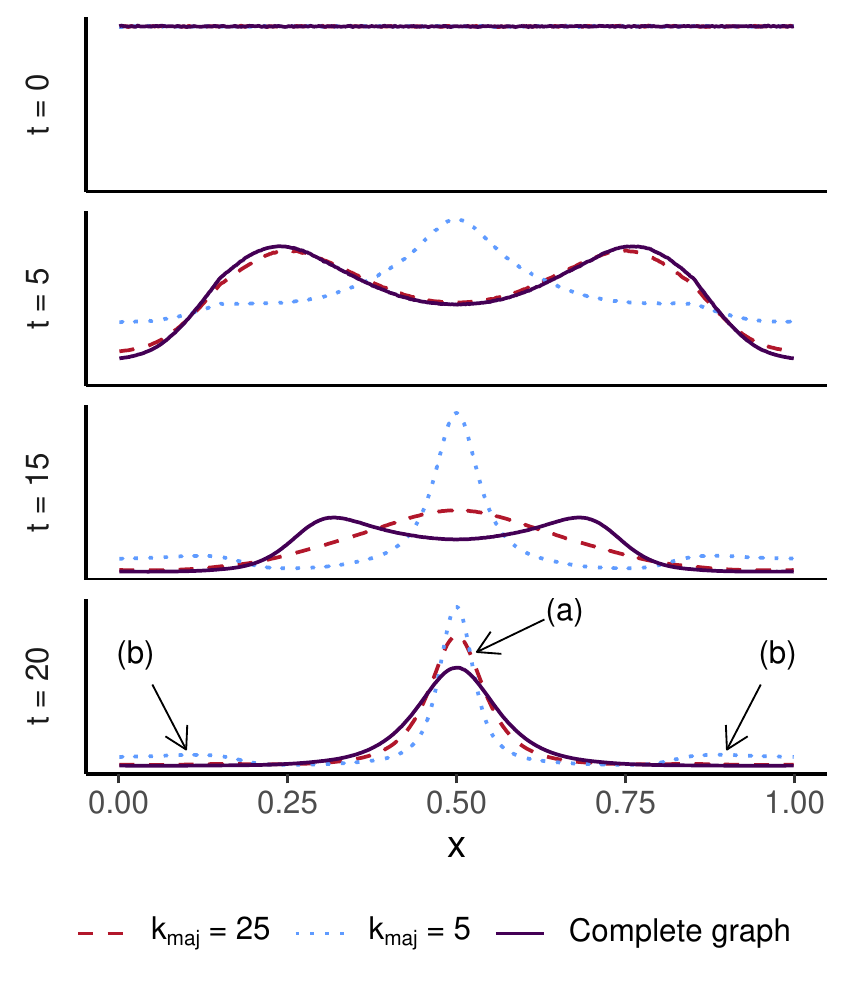}
\caption{\label{fig:dbmf_various_k}Time evolution of the opinion distribution, averaged over simulations, on configuration model networks ($N=10^4$) where $90\%$ of nodes have degree $k_{maj}$ and $10\%$ of nodes have degree 100. Snapshots are shown at the times labelled on the vertical axis (earliest time on top). Two cases are considered: $k_{maj}=25$ and $k_{maj}=5$; the latter case has lower connectivity. For comparison, the results on a complete graph (corresponding to the original MF approximation of \cite{Ben-Naim2003}) are also shown. One central cluster forms for $k_{maj} = 25$ (a), while for $k_{maj}=5$ smaller clusters also persist at the boundaries of the opinion space (b). } 
\end{figure}

\begin{figure}[!ht]
\centering
\includegraphics{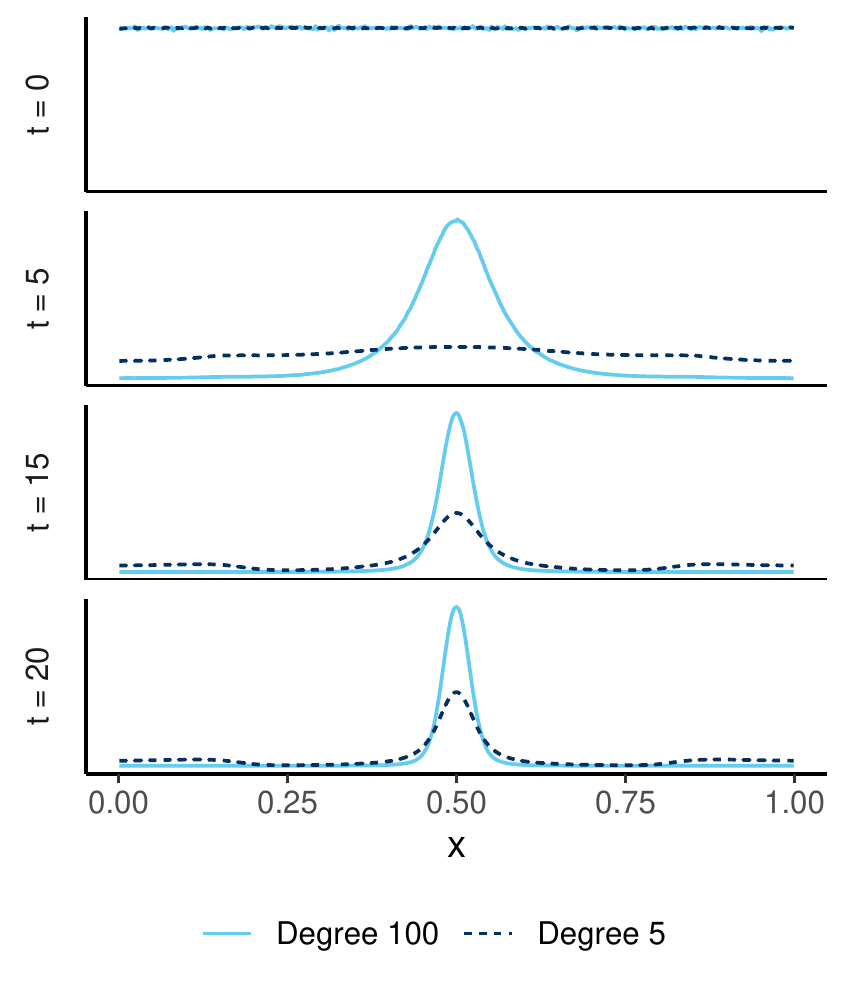}
\caption{\label{fig:dbmf5_marginal}The distribution of opinions within each degree class on a configuration model network with $90\%$ degree-5 nodes and $10\%$ degree-100 nodes.}
\end{figure}

Figure~\ref{fig:dbmf_various_k} shows the time evolution of the opinion distribution for the two cases $k_{maj} = 25$ and $k_{maj} = 5$, averaged over simulations, compared with the average opinion distribution from simulations on a complete graph. 
Firstly, note that the behaviour for the graph with $k_{maj} = 25$ is similar to that of the complete graph, with one cluster forming at $x\approx0.5$. While this large central cluster also appears in the $k_{maj} = 5$ case, we also see clusters persisting near the boundaries of the opinion space. As the distribution is quite flat at the boundaries these clusters may seem small, however they make up over $20\%$ of the network. Note that the complete-graph case does not display these boundary clusters. The distribution of opinions within each degree class is shown in Fig.~\ref{fig:dbmf5_marginal}. The higher degree nodes form one cluster at $x\approx 0.5$. For the lower degree nodes, while many are pulled towards the centre, a portion remain closer to the boundaries.

\begin{figure}[!ht]
\centering
\includegraphics{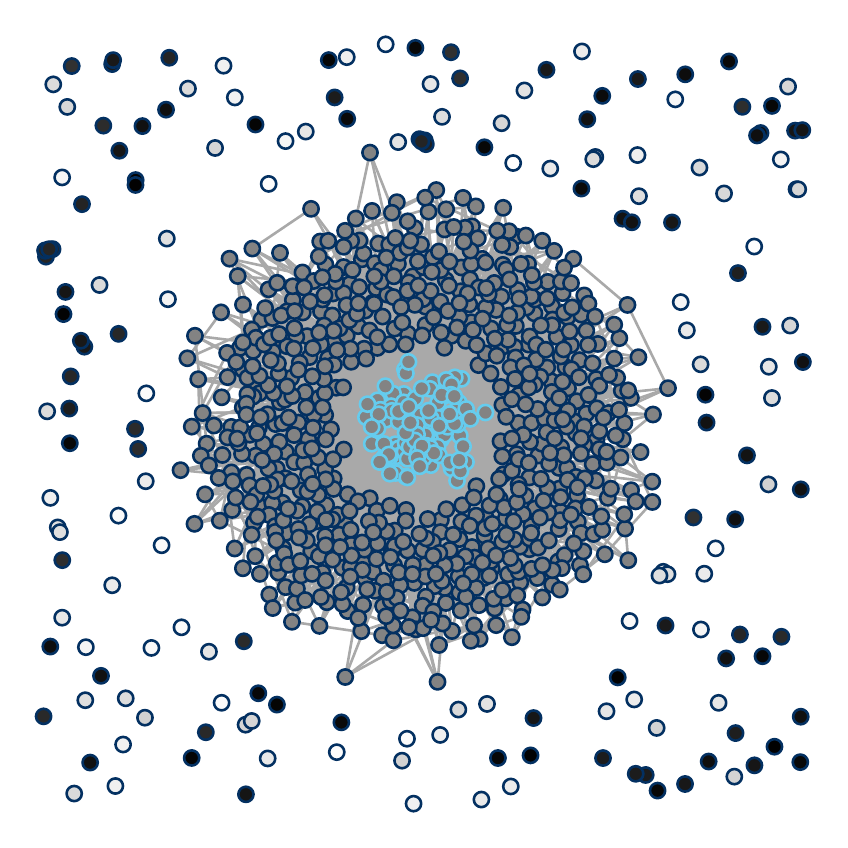}
\caption{\label{fig:network}The temporal network for one simulation on a configuration model network with $90\%$ degree 5 nodes and $10\%$ degree 100 nodes ($N=1000$) once the system has converged. Links exist between pairs of nodes who can interact, i.e., are connected in the network and have opinion difference less than $\epsilon$. $80\%$ of nodes form one cluster with opinions close to 0.5 (grey fill). The rest, in this case all degree 5 nodes (dark blue boundary), form small clusters closer to the boundaries of the opinion space (black is opinion 0, white is opinion 1).}
\end{figure}

It is useful to think of two types of network in this process: the underlying static network and a temporal network in which nodes are connected if they are both connected in the underlying network and close enough in the opinion space to influence each other. Observations of individual simulations show that nodes, typically those with low degrees, become disconnected in the temporal network as their neighbours move away from them in the opinion space. 
An example of the temporal network (with $N=1000$ for visualization purposes) once the system has converged is shown in Fig.~\ref{fig:network}. The colour of the node represents its location in the opinion space (greyscale where 0=black and 1=white). Approximately $80\%$ of nodes form a large connected component with opinions close to 0.5, whereas $20\%$ of nodes, all of which have degree 5 (indicated by the dark blue boundary), become disconnected from the large connected component. Due to the small number of connections between the low degree nodes, they are either completely isolated or form clusters with 1 or 2 other nodes. When this individual simulation is represented in terms of its opinion distribution, there is a large peak at 0.5 along with many smaller peaks at the boundaries. These smaller peaks, when averaged over many simulations, yield the uniform-like portions at the boundaries in Figs.~\ref{fig:dbmf_various_k} and~\ref{fig:dbmf5_marginal}.

\section{Degree-based mean-field equations}
\label{sec:dbmf}
To approximate the dynamics on a configuration-model network with degree distribution $q_k$ (i.e., $q_k$ is the probability that a randomly-chosen node has degree $k$), we make an annealed network assumption, which is that at each time step the edges of the network are all assumed to be rewired, while the individual nodes retain their degree. This means that the network at each time step is one of the ensemble of networks with the prescribed degree distribution  \cite{Bianconi2018}. In a given time step, the probability that a node is connected to, and therefore chosen for interaction with, a node in degree class $k$ is the same for any node in that class. 
Under this assumption we need not know the opinions of each individual node in the class, only the distribution of opinions within the class. We define the densities $P_k(x,t)$ for each distinct degree class $k$ such that (in the limit $dx\to0$) $P_k(x,t)\,dx$ is the probability that the opinion of a $k$-degree node lies in the interval $[x,x+dx)$ at time $t$.

We can write down a continuous-time equation for $P_k(x,t)$ by calculating the expected change in the number of degree $k$ nodes with opinion in $[x,x+dx)$ in an infinitesimal time step $dt$ as follows. The expected number of nodes of degree $k$ is (for large $N$) $N q_k$, and the fraction of these with opinions in $[x,x+dx)$ is $P_k(x,t)\,dx$, so the expected change over the time increment $dt$ can be written as
\begin{equation}
\begin{split}
N&q_kP_k(x,t+dt)\,dx - Nq_kP_k(x,t)\,dx =
\\ &\text{ Expected number of degree-$k$ nodes whose opinions move \emph{inside} }[x,x+dx) \text{ in $dt$} \\
&- \text{Expected number of degree-$k$ nodes  whose opinions move \emph{outside} }[x,x+dx)\text{ in $dt$},
\end{split}
\label{eqn:dbmf1}
\end{equation}
where the time step $dt = 2/N$ is chosen such that one pair of nodes interact at each time step. Note that the limit $N\to \infty$ corresponds to the continuous-time limit $dt\to0$.

To calculate the positive contribution to the right hand side of Eq.~\eqref{eqn:dbmf1}, consider a node $i$ with degree $k$ and opinion $y$. 
We define $\pi_{kl}$ to be the probability that an edge exists between a node chosen at random from all degree $k$ nodes and a node chosen at random from all degree $l$ nodes~\footnote{In terms of the adjacency matrix for the network, $\pi_{kl} = \sum_{i:deg(i)=k}\sum_{j:deg(j)=l}A_{ij}/Nq_kNq_l$ which is the ratio of the number of edges that exist between degree $k$ and degree $l$ nodes and the number of possible edges (if all pairs of nodes were connected).}.
The probability that node $i$ is chosen for interaction with a degree $l$ node in $dt$ is $\frac{Nq_l\pi_{kl}}{|E|}$, where $|E| = \frac{N^2 }{2}\sum_k\sum_lq_kq_l\pi_{kl} $ is the number of edges in the network~\footnote{On a network with no self-edges, the number of edges is $|E| = \frac{1}{2}\sum_k[\,\sum_{l\neq k}Nq_kNq_l\pi_{kl} -  Nq_k\left(Nq_k - 1\right)\pi_{kk}]$. In the limit $N\rightarrow \infty$ this is $|E| = \frac{N^2 }{2}\sum_k\sum_lq_kq_l\pi_{kl}$.}. It is useful to write this probability in terms of the graph density $\gamma =  \frac{2|E|}{N^2}$, which is the ratio of the number of edges in the network to the number of possible edges in the large $N$ limit ($\gamma = 1$ on a complete graph); the probability that node $i$ is chosen for interaction with a degree $l$ node is then $\frac{2q_l\pi_{kl}}{N\gamma}$~\footnote{Again, on a network with no self-edges the graph density is $\gamma =  \frac{2|E|}{N(N-1)}$, and the probability that node $i$ is chosen for interaction with a degree $k$ node is $\frac{2(Nq_kk-1)\pi_{kk}}{N^2\gamma}$. However the quantities defined in the text assume the limit $N\rightarrow \infty$ will be taken.}.
Node $i$ will only be influenced to change its opinion if the degree $l$ node has opinion $z\in(y-\epsilon,y+\epsilon)$, in which case node $i$ will move its opinion to $y+\mu(z-y)$. This new opinion will be in $[x,x+dx)$ for $z\in\left[y+\frac{1}{\mu}(x-y),y+\frac{1}{\mu}(x-y)+\frac{dx}{\mu}\right)$. The proportion of degree $l$ nodes with such opinions is $P_l\left(y+\frac{1}{\mu}(x-y),t\right)\frac{dx}{\mu}$. Thus, the probability that node $i$ has opinion in $[x,x+dx)$ after interacting with a degree $l$ node is $\frac{2q_l\pi_{kl}}{N\gamma}P_l\left(y+\frac{1}{\mu}(x-y),t\right)\frac{dx}{\mu}$. Summing over all degree classes, the probability that node $i$ moves its opinion to $[x,x+dx)$ in $dt$ is 
\begin{equation*}
\sum_l\frac{2q_l\pi_{kl}}{N\gamma}P_l\left(y+\frac{1}{\mu}(x-y),t\right)\frac{dx}{\mu}.
\end{equation*}
This probability is the same for any degree $k$ node with opinion $y$, of which there are $Nq_kP_k(y,t)\,dy$. Integrating over $y$, we get the expected number of degree $k$ nodes who move to opinion $[x,x+dx)$ in $dt$,
\begin{equation}
\label{eqn:pos_contr}
\sum_l\frac{2q_kq_l\pi_{kl}}{\gamma}\int_{|x-y|<\epsilon\mu}\frac{1}{\mu}P_k(y,t)P_l\left(y+\frac{1}{\mu}(x-y),t\right)\,dx\,dy,
\end{equation}
where the domain of integration comes from the constraint for $z\in(y-\epsilon,y+\epsilon)$.

Now, to calculate the negative contribution of the right hand side of Eq.~\eqref{eqn:dbmf1}, consider a node $j$ with degree $k$ and opinion in $[x,x+dx)$. Node $j$ will change its opinion if it interacts with a node with opinion $y$ such that $|x-y|<\epsilon$. The probability of this happening is
\begin{equation*}
\int_{|x-y|<\epsilon}\sum_l\frac{2q_l\pi_{kl}}{N\gamma}P_l(y,t)\,dy.
\end{equation*}
The number of nodes with degree $k$ and opinion in $[x,x+dx)$ is $Nq_kP_k(x,t)\,dx$ and so the expected number of degree $k$ nodes who move outside of $[x,x+dx)$ in $dt$ is 
\begin{equation}
\label{eqn:neg_contr}
\sum_l\frac{2q_kq_l\pi_{kl}}{\gamma}\int_{|x-y|<\epsilon}P_k(x,t)P_l(y,t)\,dy\,dx.
\end{equation}
Inserting expressions \eqref{eqn:pos_contr} and \eqref{eqn:neg_contr} into the right hand side of Eq.~\eqref{eqn:dbmf1}, we obtain
\begin{equation}
\label{eqn:dbmf2}
\begin{split}
Nq_kP_k(x,t+dt)\,dx - Nq_kP_k(x,t)\,dx = \phantom{space}& \\
\sum_l\frac{2q_kq_l\pi_{kl}}{\gamma}\left[\int_{|x-y|<\epsilon\mu}\frac{1}{\mu}\right.P_k(y,t)&P_l\left(y+\frac{1}{\mu}(x-y),t\right)\,dx\,dy\\
&-\left.\int_{|x-y|<\epsilon}P_k(x,t)P_l(y,t)\,dy\,dx\right].
\end{split}
\end{equation}
We assume the ensemble of networks is from a configuration model, in which case $\pi_{kl} = \frac{kl}{N\langle k \rangle}$, where $\langle k \rangle = \sum_k kq_k$ is the average degree~\footnote{When a configuration model network is constructed, each node $i$ is assigned a degree $k_i$ which is represented as a stub (half-edge). Two stubs are chosen at random and connected to form an edge. Then, two further stubs are chosen at random from the remaining stubs and connected. This continues until all stubs have been connected. Consider a node $i$ with degree $k$ and a node $j$ with degree $l$. If we take a stub of node $i$, the probability that this is connected to one of node $j$'s stubs is $l/2|E|$, where $2|E|$ is the total number of stubs (twice the number of edges in the network). The probability that any of node $i$'s stubs connects to any of node $j$'s, i.e., the probability that there is an edge between node $i$ and node $j$, is $kl/2|E|$. Now \unexpanded{$|E| = N\langle k \rangle/2$}, and so $\pi_{kl} = \frac{kl}{N\langle k \rangle}$.}. Rearranging Eq.~\eqref{eqn:dbmf2} and taking the limit as $dt=2/N\rightarrow 0$, assuming the number of degree classes and the degree distribution are fixed, we get 
\begin{equation}
\label{eqn:dbmf}
\begin{split}
\frac{\partial P_k(x,t)}{\partial t} = \sum_l \frac{klq_l}{\langle k \rangle^2}\left[\frac{1}{\mu} \right.\int_{|x-y|<\epsilon\mu} P_k(y,t)&P_l\left(y+\frac{1}{\mu}(x-y),t\right) \,dy \\
& \left.- \int_{|x-y|<\epsilon} P_k(x,t)P_l\left(y,t\right) \,dy \right].
\end{split}
\end{equation}

Once we have the densities within each class, we can obtain the opinion density over the network, 
\begin{equation*}
P(x,t) = \sum_k q_kP_k(x,t).
\end{equation*}

\begin{figure}[!t]
\centering
\includegraphics{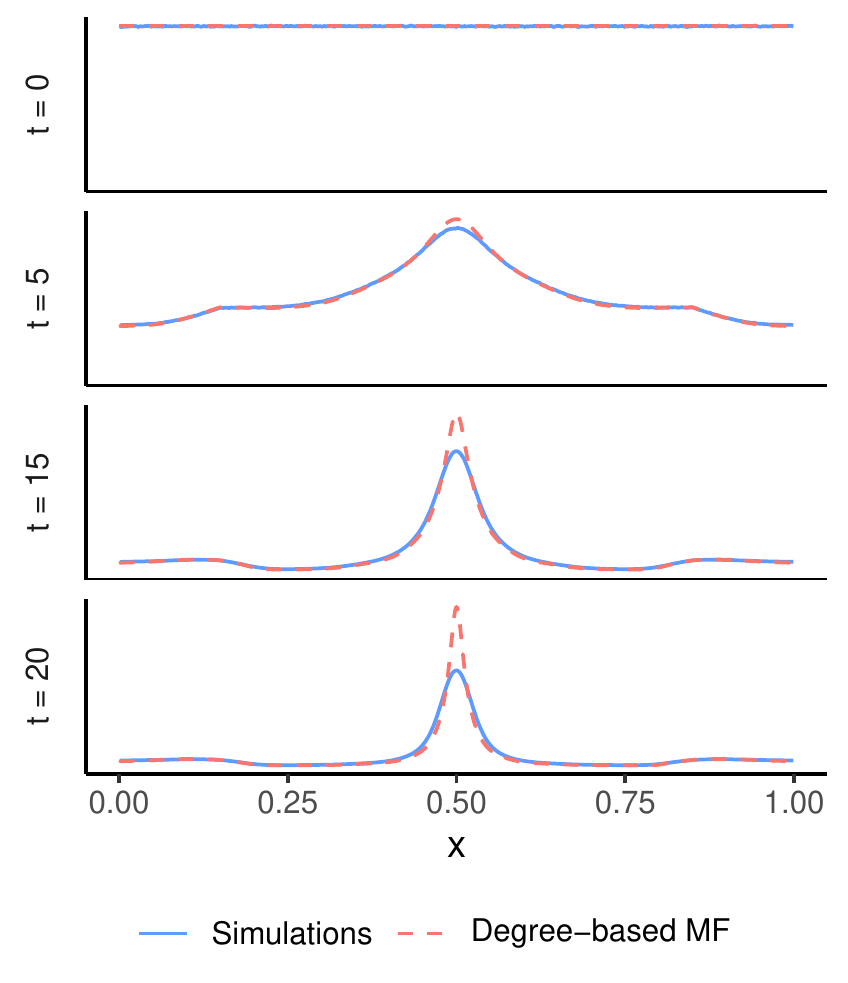}
\caption{\label{fig:dbmf_sim}The distribution of opinions on a configuration model network in which $10\%$ of nodes have degree 100 and $90\%$ of nodes have degree 5.}
\end{figure}

We solved Eq.~\eqref{eqn:dbmf} numerically for two degree classes, $k_1=5$ and $k_2=100$ $(q_{k_1}=0.9,\,q_{k_2}=0.1)$, $\epsilon = 0.3$ and $\mu=0.5$ by discretising the opinion space into 800 equally spaced states~\footnote{We checked that the results are visually unchanged if a larger number of states is used.}. The total distribution $P(x,t)$ is shown in Fig.~\ref{fig:dbmf_sim} together with the distribution from the simulations as described in section \ref{sec:effect_network}. The degree-based MF equations provide a good approximation to the simulations, and in particular they correctly predict the behaviour at the extremes of the opinion space where many of the degree 5 nodes persist. The two distributions begin to diverge at later times as the distribution from the degree-based MF equations (which assumes the $N\to \infty$ limit) evolves towards a delta-function spike at $x=0.5$, while the finite-size network in simulations retains some granularity in opinions. This is expected, however, since the amplitude of any peak in the distribution from the simulations will be limited by the network size (the effect of network size on simulation results is  discussed at the end of section~\ref{sec:sbm}).

\section{Class-based mean-field equations}
\label{sec:cbmf}
One of the key assumptions in deriving the degree-based MF equations is that there is the same probability of connecting any degree $k$ node to any degree $l$ node. However, there are many networks where this assumption is violated.
For example, in a network with community structure the probability of a degree $k$ node being connected to a degree $l$ node will be higher if the two nodes are in the same community than if they are in different communities. 

We can extend the degree-based MF equations by partitioning the network into more general classes. 
Suppose we have a network with $N$ nodes which can be partitioned into $K$ classes with a proportion $q_k$ of nodes in class $k$. We assume that this network comes from an ensemble of networks such that when we choose a network at random, and two nodes at random from that network, the probability that they are connected will depend only on the classes to which they belong. We can then take the same approach as in section \ref{sec:dbmf}, where $\pi_{kl}$ is now the probability that an edge exists between a node chosen at random from class $k$ and a node chosen at random from class $l$ and $P_k(x,t)$ is the opinion density for nodes in class $k$. 

The derivation for general classes is the same as for degree-classes up until the point where we specify $\pi_{kl}$, and so Eq.~\eqref{eqn:dbmf2} holds for this more general $\pi_{kl}$. We assume the normalised edge probabilities $\frac{\pi_{kl}}{\sum_m\sum_n\pi_{mn}}$ do not depend on $N$, which allows us to take the limit $dt = 2/N \rightarrow 0$ in Eq.~\eqref{eqn:dbmf2} to obtain
\begin{equation}
\label{eqn:cbmf}
\begin{split}
\frac{\partial}{\partial t}P_k(x,t) = \sum_l\frac{q_l\pi_{kl}}{\gamma}\left[\frac{1}{\mu}\right.\int_{|x-y|<\epsilon\mu}P_k(y,t)&P_l\left(y+\frac{1}{\mu}(x-y),t\right)\,dy \\
&-\left.\int_{|x-y|<\epsilon}P_k(x,t)P_l(y,t)\,dy\right].
\end{split}
\end{equation}
The original MF equation \cite{Ben-Naim2003} can be recovered from Eq.~\eqref{eqn:cbmf} with 1 degree class ($q_k = 1$) and $\mu=1/2$. In fact this is independent of the edge probability $\pi_{kk}$, which indicates that the dynamics are the same on complete graphs ($\pi_{kk} = 1$) and Erd\H{o}s R\'{e}nyi graphs ($\pi_{kk} = p$) for any edge probability $p$, as has previously been shown in simulation studies \cite{Meng2018}.

\section{Dynamics on a network with community structure}
\label{sec:sbm}

In a network with community structure there are, on average, more connections between nodes in the same community than between nodes in different communities. If there is greater variance in the initial opinions across communities than within them, we might expect nodes within the same community to become closer in the opinion space while nodes in different communities become separated from each other. It is natural then to use the class-based MF equations to approximate the dynamics on these networks.

\begin{figure}
\centering
\includegraphics{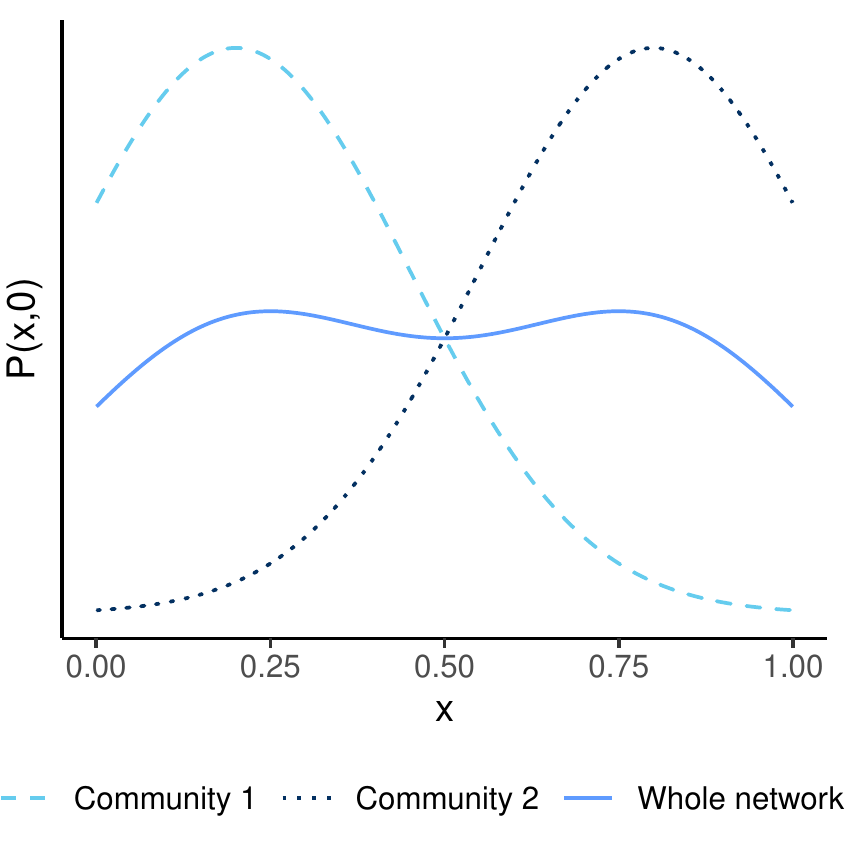}
\caption{\label{fig:initial_dist}The initial distribution of opinions within each community and over the whole network for the simulations on SBM networks.}
\end{figure}

The stochastic block model (SBM) is a random graph model which can produce networks with community structure. Nodes are assigned to groups and each pair of nodes is connected with a probability dependent on their group membership. If this probability is higher for nodes in the same group than nodes in different groups then the network will have a community structure. We simulate the dynamics on networks generated from this model with $N=10^4$ nodes and two equally sized communities, where the probability of an edge between two nodes in the same community is $p_w=0.1$ and the probability of the edge between two nodes in different communities is $p_b=0.01$. Each node is assigned an initial opinion from a truncated normal distribution with mean 0.2 or 0.8, depending on which community that node belongs to, and standard deviation 0.25. The initial distribution in each community, and the distribution over the whole population, is shown in Fig.~\ref{fig:initial_dist}. As in the example of section \ref{sec:twodegree}, $\epsilon = 0.3$ and $\mu = 0.5$. We also simulate the dynamics on a complete graph with the same initial distribution of opinions.

\begin{figure}[!ht]
\centering
\includegraphics{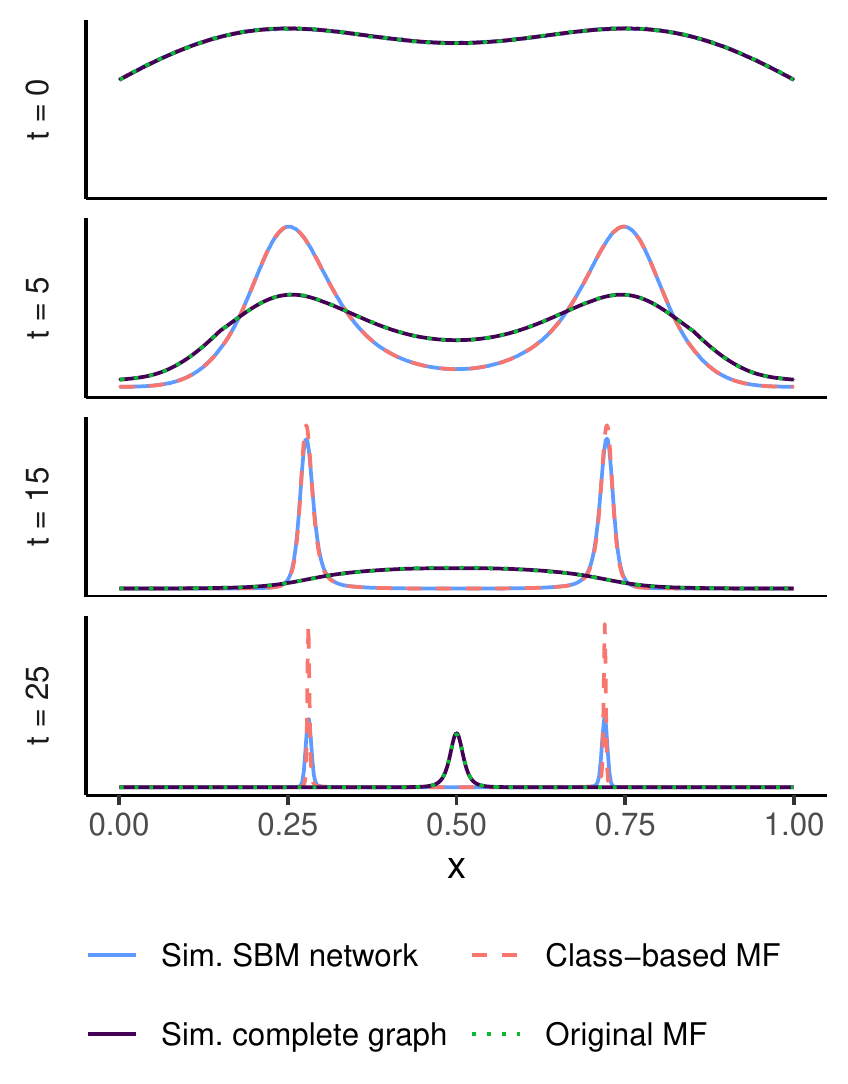}
\caption{\label{fig:sbm_pw01_pb001_tot}Opinion distribution, averaged over simulations, on a SBM network with $p_w = 0.1$ and $p_b=0.01$ and on a complete graph, together with the distributions from the two MF approximations.}
\end{figure}

The opinion distribution  at a number of time points is shown in Fig.~\ref{fig:sbm_pw01_pb001_tot}. On the SBM network the two communities become separated in the opinion space, while on a complete graph consensus emerges. The distribution calculated from the class-based MF equations is a good match to the distribution from the simulations. As we would expect, the original MF equation predicts the central cluster that we see on the complete graph. 

\begin{figure}[!ht]
\centering
\includegraphics{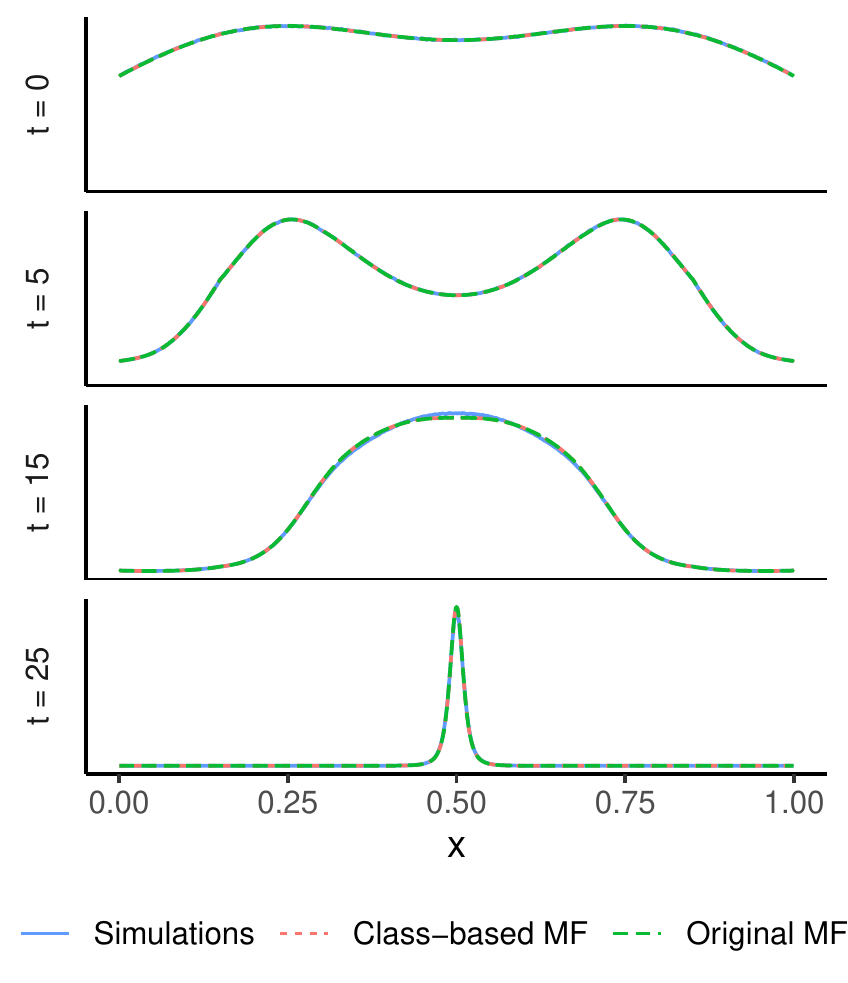}
\caption{\label{fig:sbm_pw01_pb01_tot}Opinion distribution on a SBM network with $p_w = 0.1$ and $p_b=0.1$.}
\end{figure}

Not all networks generated from the stochastic block model have community structure. If $p_b=p_w$ then there are as many connections between communities as within communities. Since nodes can be influenced as much from agents outside of their community as from those within it, this leads to consensus throughout the network as can be seen in Fig.~\ref{fig:sbm_pw01_pb01_tot} for $p_b=p_w=0.1$. In this case we do not need a class-based approach to approximate the dynamics as the original MF equation is sufficient, and in fact for $p_b=p_w$ the class-based MF equations reduce to the original MF equation. 

\begin{figure}[!ht]
\centering
\includegraphics{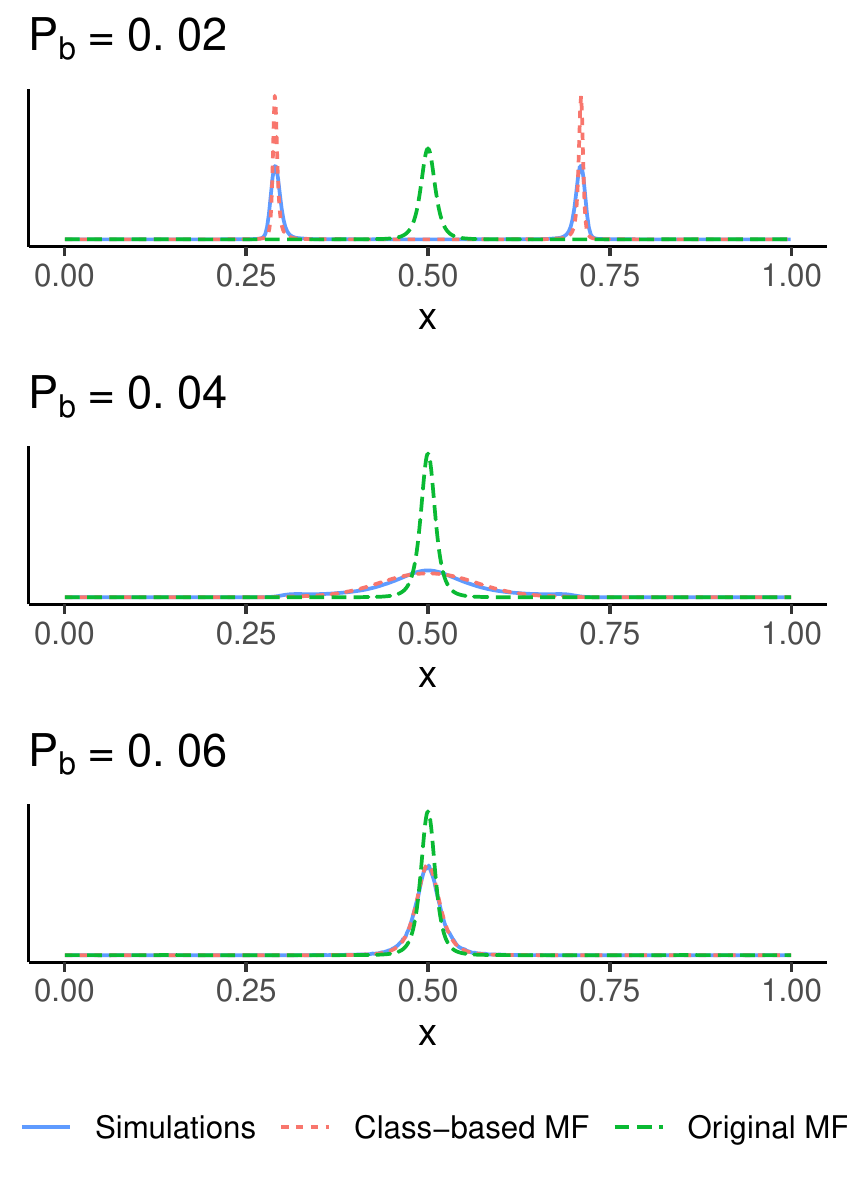}
\caption{\label{fig:sbm_various_p}Opinion distributions on SBM networks for $p_b = 0.02,\,0.04$ and $0.06$ with $p_w=0.1$, all at $t=25$.}
\end{figure}

Even when $p_b\neq p_w$ the community structure may not be very pronounced and the original MF equations might be sufficient to describe the dynamics. Figure~\ref{fig:sbm_various_p} shows the distributions from simulations on SBM networks for $p_w=0.1$ and $p_b = 0.02,\, 0.04$ and $0.06$. For $p_b = 0.02$ a class-based approach is needed to pick up the two clusters that form. For larger values of $p_b$ the original MF equation correctly predicts the qualitative behaviour, although as we would expect it does not provide as good a fit to the distribution as the class-based MF does. 

This behaviour can to some extent be predicted from a closer inspection of Eq.~\eqref{eqn:cbmf}. The strength of the coupling between the opinion distributions is controlled by the ratio of the coefficients within an equation, $\frac{q_2\pi_{12}/\gamma}{q_1\pi_{11}/\gamma}$. The two communities are of equal size, so this ratio is $\frac{p_b}{p_w}$. For $p_b = 0$ the distributions become uncoupled and we get two systems evolving independently in the original MF regime, resulting in polarization. For $p_b = p_w$ the system of equations can be reduced to the original MF equation with ${P(x,t) = \frac{1}{2}(P_1(x,t)+P_2(x,t))}$ and in this setting we get consensus.  At an intermediate value $p_b^*\approx 0.0275$ a bifurcation occurs. For $p_b<p_b^*$ the coupling is not strong enough to prevent polarization, while for $p_b>p_b^*$ the coupling is strong enough to result in consensus.

\begin{figure}[ht!]
\centering
\includegraphics{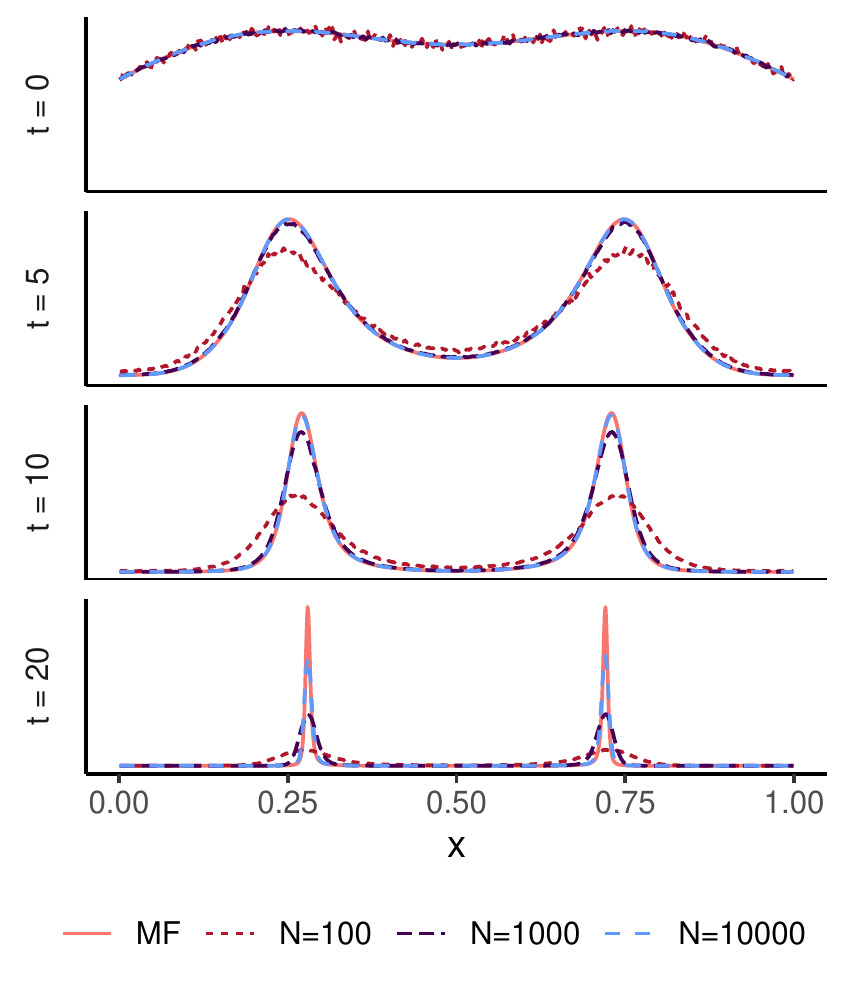}
\caption{\label{fig:effect_N}Opinion distributions on SBM networks with $p_w=0.1$ and $p_b = 0.01$ for different network sizes and from the class-based MF approximation.}
\end{figure}

For most of the examples we have shown, the distribution from the simulations clearly starts to diverge from the MF distribution at later times. This is because we are approximating the dynamics on a finite sized network using an equation that is valid in the limit of large network size. Figure~\ref{fig:effect_N} shows that as we increase the size of the network, the length of time for which the MF equations accurately predict the simulated distribution  increases. This effect will be more apparent in section~\ref{sec:real_networks} when we simulate the dynamics on real-world networks which are somewhat smaller than those we have simulated; the key point, however, is that the proposed MF approximation accurately predicts the locations of the clusters.

\section{Examples on real networks}
\label{sec:real_networks}

In the previous section we showed that the class-based MF is needed to accurately approximate the opinion distribution on networks with community structure. However, our examples were limited to large  synthetic networks. 
We now compare the class-based MF approximation with simulations on three real-world networks with community structure: Zachary's Karate Club network \cite{Zachary1977}, a network of books on politics sold by Amazon.com \cite{Krebs} and two co-voting networks from the U.S. House of Representatives \cite{voteview,Waugh2012}. The networks chosen have ground-truth communities which offer a natural partition of the network into classes. 

For each network we average over $10^4$ Monte Carlo simulations. Each node takes an initial opinion from a truncated normal distribution with a mean that depends on the class of the node and a standard deviation of 0.25. We calculate the edge probabilities for the class-based MF equations by summing the number of edges between the two classes and dividing by the number of possible edges ($N^2q_kq_l$ for different classes, $Nq_k(Nq_k-1)$ for the same class).

\subsection{Zachary's Karate Club}
\begin{figure}[ht!]
\centering
\includegraphics{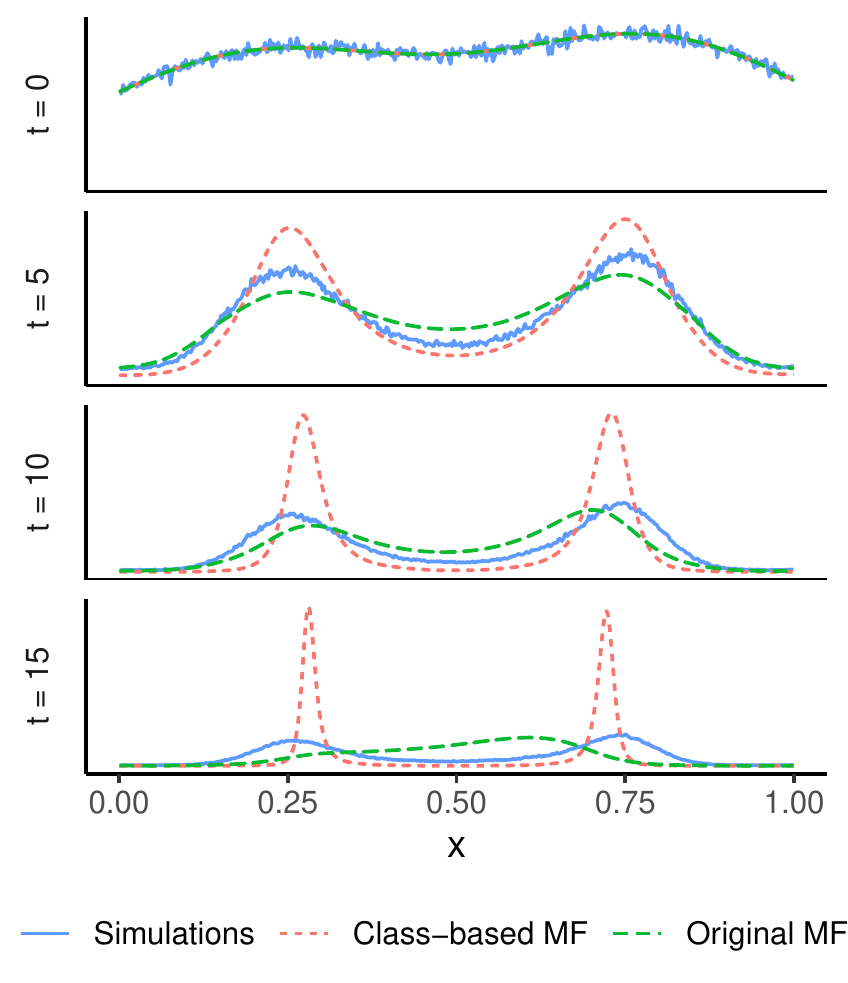}
\caption{\label{fig:karate_tot}Distribution of opinions on Zachary's karate club network.}
\end{figure}

Zachary's Karate Club network \cite{Zachary1977} documents the friendships of 34 members of a karate club. Two communities correspond to a split in the club after a disagreement between the administrator of the club and the club's instructor. We assign initial opinions to community one (two) from a truncated normal distribution with mean 0.2 (0.8) and standard deviation 0.25.  Simulations of the dynamics on this network show that two clusters form in the opinion space, which correspond to the two communities, see Fig.~\ref{fig:karate_tot}. The class-based MF equations capture these clusters, although the opinion profile starts to diverge from that of the simulations quite early since the network has only 34 nodes. In contrast, the original MF equation incorrectly predicts consensus on this network.

\subsection{Politics books}
\begin{figure}[!t]
\centering
\includegraphics{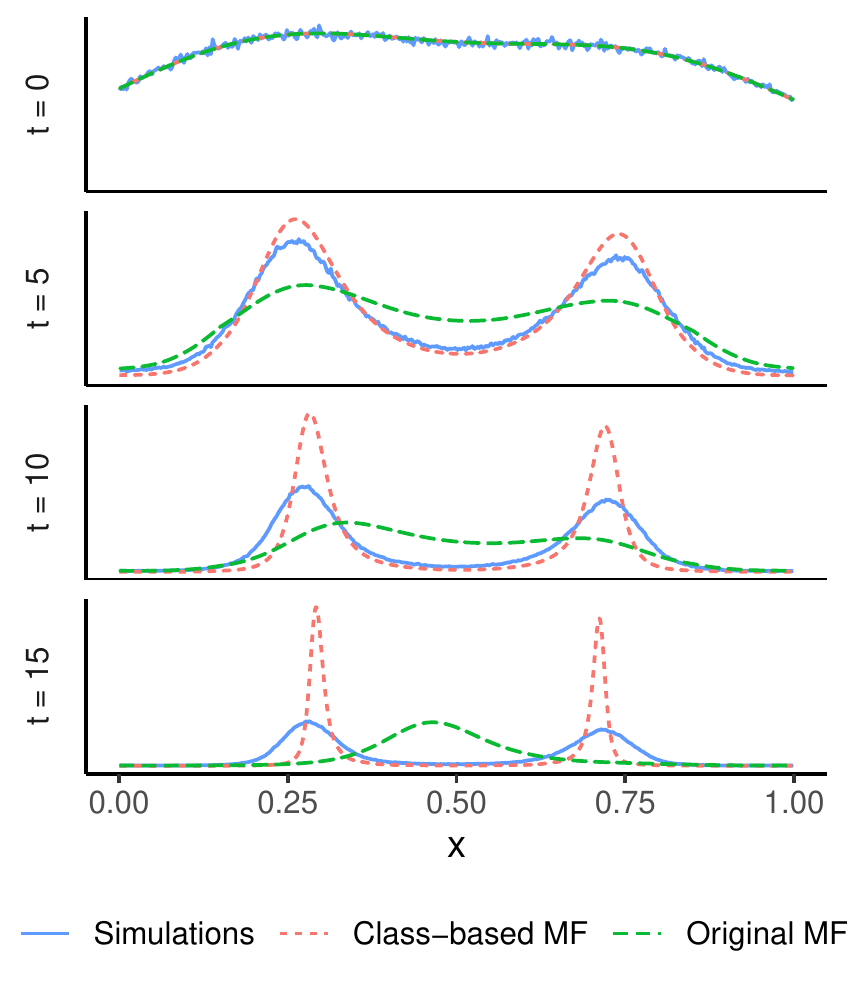}
\caption{\label{fig:polbooks_tot}Distribution of opinions on the politics books network.}
\end{figure}

We next look at a network of 105 books on US  politics \cite{Krebs} which were sold by Amazon.com. Edges exist between books which were frequently co-purchased by the same buyers. 
The network is partitioned into communities based on the book's classification as `liberal', `neutral' or `conservative'; we assign a mean of 0.2, 0.5 and 0.8 respectively to the initial distribution for each community. In this case, much like the karate club case, the class-based MF equations predict the two clusters as per the simulations, while the standard MF equation predicts consensus as shown in Fig.~\ref{fig:polbooks_tot}. 

\begin{figure}[!htb]
\centering
\includegraphics{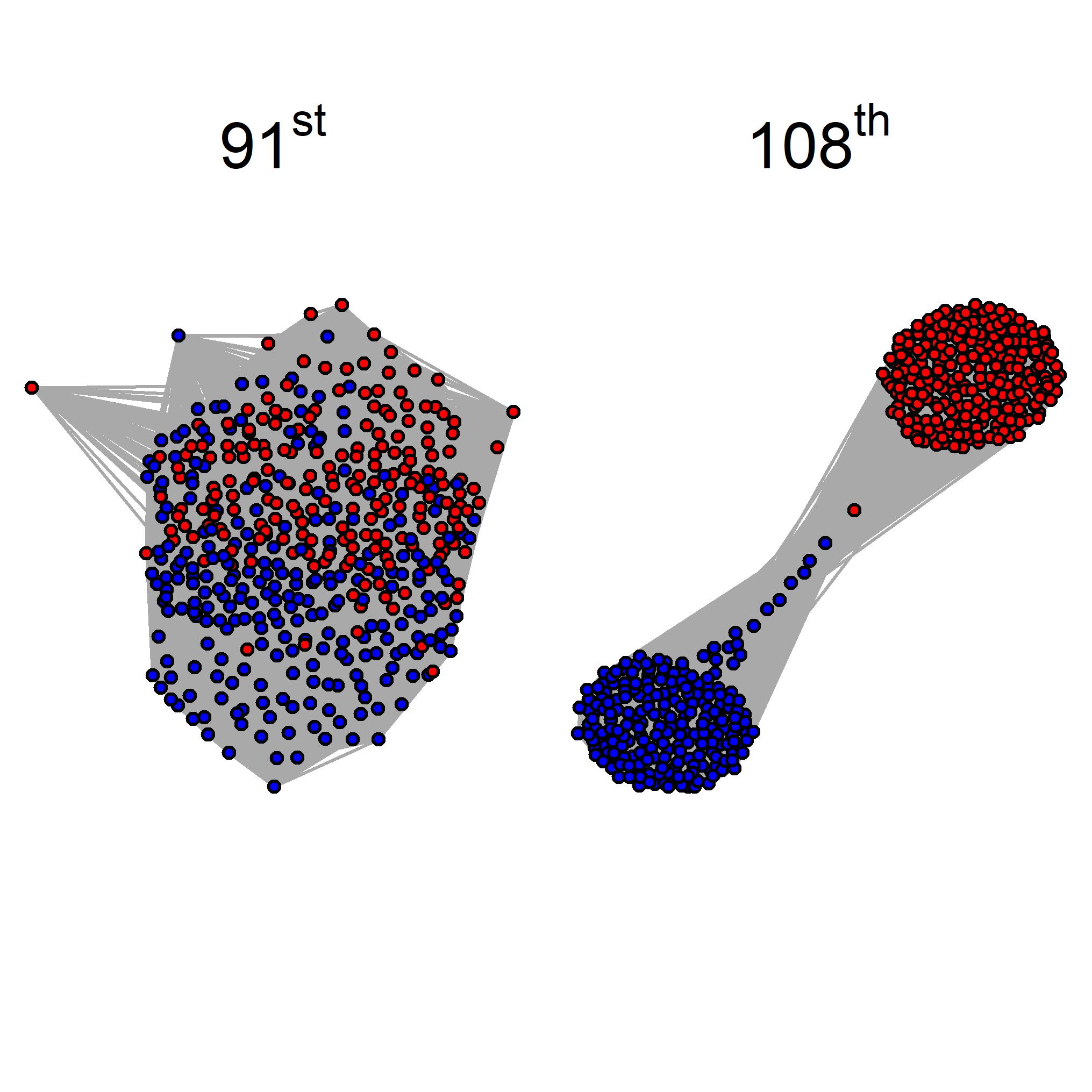}
\caption{\label{fig:networks_hse}Co-voting networks in the 91st and 108th U.S. House of Representatives. Edges exist between members who voted the same way on at least $50\%$ of bills.}
\end{figure}

\subsection{Co-voting networks}
Finally, we consider two co-voting networks from the US House of Representatives, the 91st (1969-1971) and the 108th (2003-2005) \cite{voteview,Waugh2012}. 
Edges exist between members who voted the same way on more than 50\% of bills, and the networks are partitioned based on party membership. 
We chose the 108th House as this had one of the highest ratios of within-party edges to between-party edges, and the 91st as it had one of the lowest. 
These networks are shown in Fig.~\ref{fig:networks_hse}. Following our other studies, we assign initial mean opinions according to the group membership with 0.2 for Democrats and 0.8 for Republicans.

\begin{figure}[!t]
\centering
\includegraphics{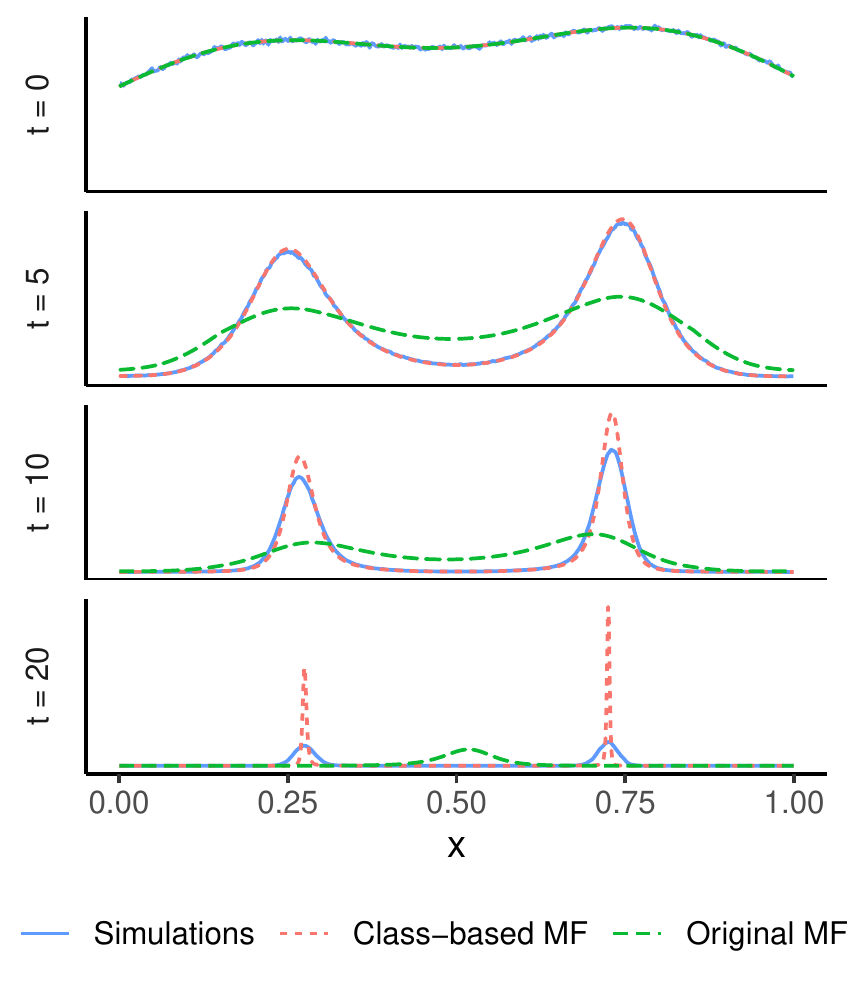}
\caption{\label{fig:house108_tot}Distribution of opinions on the 108th House of Representatives co-voting network.}
\end{figure}

\begin{figure}[ht!]
\centering
\includegraphics{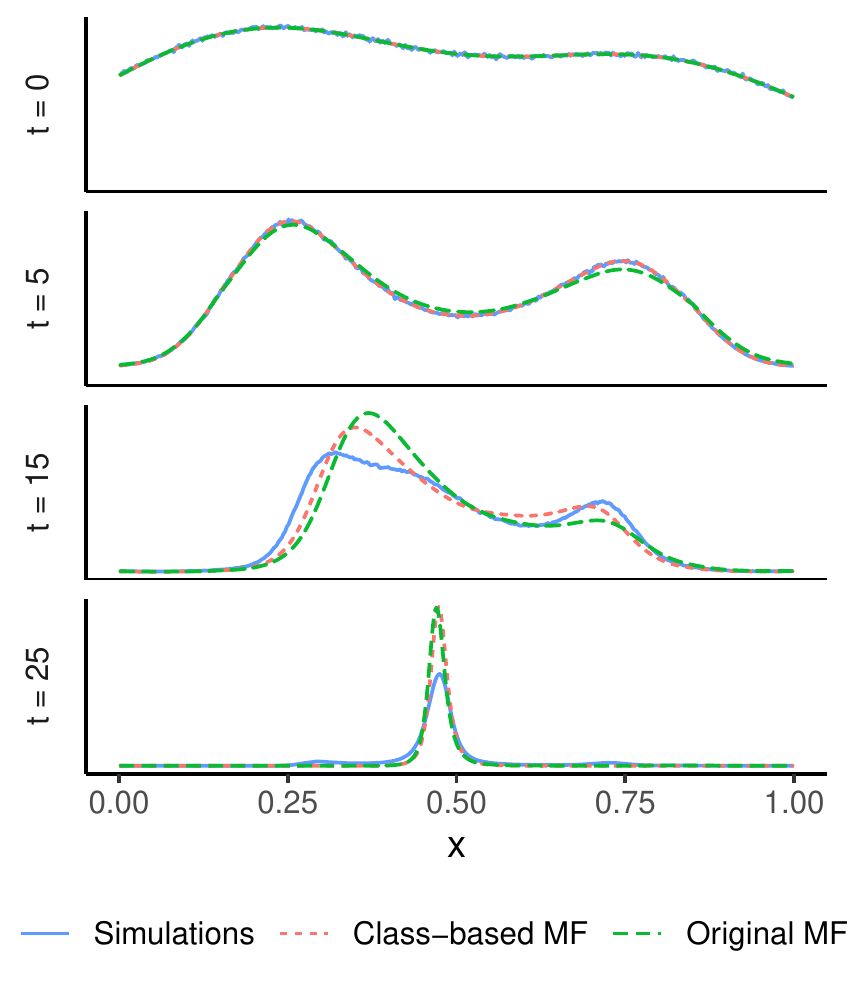}
\caption{\label{fig:house91_tot}Distribution of opinions on the 91st House of Representatives co-voting network.}
\end{figure}

For the 108th House network we obtain similar results to those of the karate club and the politics books networks, with two clusters forming in the opinion space, see Fig.~\ref{fig:house108_tot}. Note, however, that the class-based MF approximation matches the simulation results for longer here than in the previous two cases, which is expected for a larger network. Fig~\ref{fig:house91_tot} shows that on the network for the 91st House, which has very low modularity, consensus is formed. As there are almost as many edges between groups as there are within groups, the dynamics evolve on this network in a similar way to the dynamics on a complete graph. While the class-based MF equations provide a slightly better approximation than the standard MF equation, both approaches match the simulation results quite well.


\subsection{Summary of real networks}

\begin{figure}[ht!]
\centering
\includegraphics{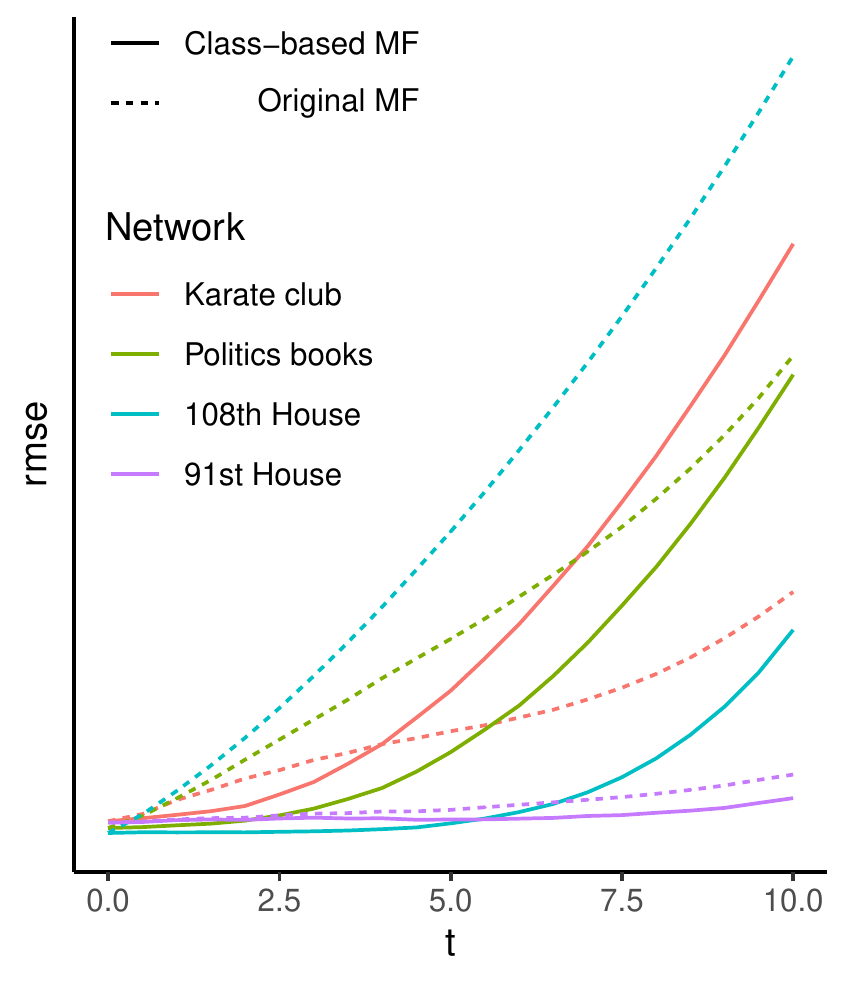}
\caption{\label{fig:rmse}Root mean square error between the density from the simulations and the density from the MF approximation.}
\end{figure}

The class-based MF equations accurately predict the qualitative behaviour of the dynamics on the four real-world networks we have studied. They are also quantitatively very accurate, although as expected this accuracy decreases at later times due to finite-size effects. 
Figure~\ref{fig:rmse} shows the root mean square error (RMSE) as a measure of distance between the density from the simulations and the density from the MF equations for each network over time. 
The error is much lower for the class-based MF approximation than the original MF approximation on the networks with higher modularity, echoing what we saw earlier. The difference in error is much lower on the 91st House of Representatives network since this has very low modularity and the original MF is already a good approximation to the dynamics in this case. 

On the Karate Club network the RMSE for the class-based MF begins to exceed the RMSE for the original MF at $t\approx 3.5$. If we take the difference in RMSE ($\text{class-based MF}-\text{original MF}$) then we see negative values at later times, as Table~\ref{table:network_metrics} shows. However, we also see these negative values on the SBM network with $N=100$ nodes. As previously mentioned, the quantitative behaviour of the MF approximation is only accurate up to a certain time, which depends on the network size, and this should be kept in mind when using metrics that directly compare the opinion distributions. At later times the qualitative behaviour should be considered instead, and it is clear from Fig.~\ref{fig:karate_tot} that the class-based MF approximation  correctly predicts the emergence of two distinct clusters in the opinion space, which the original MF approximation fails to do.

\begin{table}
\centering
\caption{\label{table:network_metrics}Network size, modularity and performance metrics for the MF approximations  on 4 real world networks and SBM networks with $p_w = 0.1$. The modularity for each SBM network is the average over $10^4$ simulations. The difference in RMSE is the RMSE for the class-based MF minus the RMSE for the original MF. Positive values indicate that the proposed class-based MF was a better approximation to the simulation distribution that the original MF.}
\begin{tabular*}{\textwidth}{c@{\extracolsep{\fill}}ccccccc}
\hline \hline
\multicolumn{2}{c}{\multirow{2}{*}{Network}} & Number & \multirow{2}{*}{Modularity}&\multicolumn{2}{c}{Difference in} &\multicolumn{2}{c}{MF correctly predicts }\\ 
&&of nodes&&\multicolumn{2}{c}{RMSE}&\multicolumn{2}{c}{consensus/polarization}\\   \hline
&&&& t = 3 & t = 10  &Class-based & Original  \\\cline{5-6}\cline{7-8}
  \multicolumn{2}{c}{Karate club}   & 34 & 0.37 & 0.04 & -0.68  &Yes&No\\ 
  \multicolumn{2}{c}{Politics books}   & 105 & 0.41 & 0.17 & 0.04  &Yes&No\\ 
  \multicolumn{2}{c}{108th House}   & 440 & 0.47 & 0.30 & 1.12  &Yes&No\\ 
  \multicolumn{2}{c}{91st House}   & 448 & 0.05 & 0.01 & 0.05  &Yes&Yes\\ 
  \hline
  \multicolumn{2}{c}{SBM\phantom{sp}}&&&&&&\\\cline{1-2}
  \multirow{10}{*}{$p_b = $} & 0.01 & \multirow{10}{*}{$10^4$} & 0.41 & 0.26 & 1.41  &Yes&No\\ 
   & 0.02 &  & 0.33 & 0.21 & 1.08  &Yes&No\\ 
   & 0.03 &  & 0.27 & 0.16 & 0.81  &Yes&Yes\\ 
   & 0.04 &  & 0.21 & 0.12 & 0.61  &Yes&Yes\\ 
   & 0.05 &  & 0.17 & 0.09 & 0.44  &Yes&Yes\\ 
   & 0.06 &  & 0.13 & 0.07 & 0.31  &Yes&Yes\\ 
   & 0.07 &  & 0.09 & 0.04 & 0.20  &Yes&Yes\\ 
   & 0.08 &  & 0.06 & 0.02 & 0.11  &Yes&Yes\\ 
   & 0.09 &  & 0.03 & 0.01 & 0.04  &Yes&Yes\\ 
   & 0.1 &  & $<0.01$ & 0.00 & 0.00  &Yes&Yes\\ \noalign{\vskip 3mm}
   \multicolumn{2}{c}{\multirow{3}{*}{$p_b = 0.01$}}  & $10^2$ & 0.41 & 0.10 & -0.33  &Yes&No\\ 
   &  & $10^3$ & 0.41 & 0.25 & 1.10  &Yes&No\\ 
   &  & $10^4$ & 0.41 & 0.26 & 1.41  &Yes&No\\ 
   \hline \hline
\end{tabular*}
\end{table}

\section{Conclusions}
\label{sec:conclusions}
In this paper we have considered mean-field approximations for the Deffuant opinion dynamics on networks. We derived a system of equations for the evolution of degree-based density functions and showed that the solution of these equations improves over the original, fully-mixed, mean-field approximation of \cite{Ben-Naim2003} in cases where the network is sufficiently different from a  complete graph. We extended the scope of the MF approximation by generalizing from degree-based classes to arbitrary class labels, and demonstrated the utility of this method for describing dynamics on networks with community structure. 

One limitation of the MF approximation is that it is derived under the assumption of infinite network size (i.e., in the limit $N\to\infty$). Indeed, our comparisons with Monte Carlo simulations show that the accuracy of the approximation is better for large networks than for small ones, see Fig.~\ref{fig:effect_N}. Despite this, and other limitations of MF approximations on real-world networks \cite{Porter2016}, we find reasonable qualitative agreement between the predictions of the MF equations and Monte Carlo simulations on real-world networks of modest size ($N$ ranging from 34 to 448). The quantitative agreement is reduced at later times as the finite-size effects cause the long-term opinion density to differ from that of the MF approximation.

A MF approach is useful because it is more computationally efficient than Monte Carlo simulation of dynamics, particularly on large networks. We can also gain some insight from a mathematical analysis of the MF equations, and simply writing down the equations for a given network can afford some understanding of the dynamics.
On an Erd\H{o}s R\'{e}nyi graph, for example, the class-based MF equations reduce to the original MF equation, indicating that the dynamics on these networks are the same as on a complete graph. On stochastic block model networks, we can deduce that a bifurcation occurs at some value of $\frac{p_b}{p_w}$, the ratio of between community edges to within community edges. Below this value the network becomes polarized while above it consensus is reached. 

The success of the MF approximation for Deffuant dynamics opens the possibility of further extensions of the methods used here. 
An abundance of opinion dynamics models exist, and understanding how these models differ is an important challenge in computational social science \cite{Flache2017}. Even the Deffuant model has several variants on networks. Take, for example, the selection of the interacting agents in each timestep: in our work, as in the original formulation of the model, an edge is chosen at random from the network \cite{Deffuant2000}, while in other studies an agent is chosen at random together with one of their neighbours \cite{Weisbuch2004,Stauffer2004a}. Applying a similar approach as used here to derive the  generalised MF approximation for other cases could aid in understanding how these apparently small changes in the model specifications  impact the dynamics. 

We hope that this work has demonstrated the potential usefulness of generalized MF approximations for dynamics with continuous-valued nodal variables, and we anticipate further extensions of the approximation scheme to a range of dynamics on networks.

\section*{Acknowledgements}
This work is partly supported by the Irish Research Council (S.F.), by Science Foundation Ireland (grant numbers 16/IA/4470, 16/RC/3918, 12/RC/2289 P2 and  18/CRT/6049) with co-funding from the European Regional Development Fund (J.G.) and by the European Research Council (ERC) under the European Union's Horizon 2020 research and innovation programme (M.Q., grant agreement No. 802421).

\bibliographystyle{plain}
\bibliography{Deffuant_paper}

\end{document}